\documentclass[aps,twocolumn,showpacs]{revtex4}
\usepackage{amsmath}
\usepackage{epsfig}

\begin{document}

\title{Investigating tetraquarks composed of $us\bar{d}\bar{b}$ and $ud\bar{s}\bar{b}$}

\newcommand*{\NJNU}{Department of Physics, Nanjing Normal University, Nanjing, Jiangsu 210097, China}\affiliation{\NJNU}

\author{Hongxia Huang}\email{hxhuang@njnu.edu.cn}\affiliation{\NJNU}
\author{Jialun Ping}\email{jlping@njnu.edu.cn}\affiliation{\NJNU}

\begin{abstract}
In the framework of the quark delocalization color screening model, we investigate tetraquarks composed of $us\bar{d}\bar{b}$ and $ud\bar{s}\bar{b}$ in two structures: meson-meson structure and diquark-antidiquark structure.
Neither bound state nor resonance state is found in the system composed of $us\bar{d}\bar{b}$. The reported $X(5568)$ cannot be explained as a molecular state or a diquark-antidiquark resonance of $us\bar{d}\bar{b}$ in present calculation. However, two bound states of the diquark-antidiquark structure are obtained in the tetraquarks system composed of $ud\bar{s}\bar{b}$: an $IJ=00$ state with the mass of $5701$ MeV, and an $IJ=01$ state with the mass of $5756$ MeV, which maybe the partners of $X(5568)$ states. Our results indicate that the diquark-antidiquark configuration would be a good choice for the tetraquarks $ud\bar{s}\bar{b}$ with quantum numbers $IJ=00$ and $IJ=01$. The tetraquarks composed of $ud\bar{s}\bar{b}$ is more possible to form bound states than the one composed of $us\bar{d}\bar{b}$. These bound states are worth investigating in future experiments.
\end{abstract}


\maketitle

\setcounter{totalnumber}{5}

\section{\label{sec:introduction}Introduction}
In the past few decades, the discovery of numbers of exotic states stimulated extensive interest in understanding
the structures of the multiquark hadrons. So far, most tetraquark and pentaquark candidates are composed of hidden
charm or bottom quarks. However, the new state $X(5568)$ observed by the D0 collaboration in 2016~\cite{D0} was an
exception. The $X(5568)$ has a mass $m=5567.8 \pm 2.9(stat)^{+0.9}_{-1.9}(syst)$ MeV and width $\Gamma=21.9 \pm 6.4(stat)^{+5.0}_{-2.5}(syst)$ MeV~\cite{D0}. The decay mode is $X(5568) \rightarrow B^{0}_{s}\pi^{\pm}$, which
indicates that the quark component of the $X(5568)$ should be four different flavors: $u, d, s, b$.
Therefore, the claimed $X(5568)$, if confirmed, would differ from any of the previous observations, as it must be
a tetraquark state with $us\bar{d}\bar{b}$ or $ds\bar{u}\bar{b}$ and their charge-conjugated ones. Unfortunately,
this state was not confirmed by other collaborations. The LHCb collaboration~\cite{LHCb}, the CMS collaboration of
LHC~\cite{CMS}, the CDF collaboration of Fermilab~\cite{CDF} and the ATLAS Collaboration of LHC~\cite{ATLAS} all
claimed that no evidence for this state was found. Nevertheless, the D0 collaboration's new result still insists
on the existence of this tetraquark $X(5568)$~\cite{D02}. Clearly, more other measurements are needed.

The discovery of this exotic state $X(5568)$ also stimulates the theoretical interest. Many approaches have been
applied to interpret this state, such as the QCD sum rules~\cite{Agaev,WangZG1,WangZG2,Zanetti,ChenW,Dias,TangL},
quark models~\cite{WangW,XiaoCJ,Stancu}, the extended light front model~\cite{KeHW}, rescattering effects~\cite{LiuXH},
and so on. However, several theoretical calculations gave the negative results~\cite{ChenXY,Burns,GuoFK,Albaladejo}.
For example, in Ref.~\cite{ChenXY}, authors investigated two structures, diquark-antidiquark and meson-meson,
with all possible color configurations by using the Gaussian expansion method, and they cannot obtain the reported
$X(5568)$. Ref.~\cite{Burns} examined the various interpretations of the state $X(5568)$ and found that the threshold,
cusp, molecular, and tetraquark models were all unfavored the existence of the $X(5568)$.

To search for the tetraquark states with four different flavors, a better state is $ud\bar{s}\bar{b}$ (or its
charge-conjugated one) with replacing the $d/\bar{s}$ in $X(5568)$ by $\bar{d}/s$~\cite{YuFS}. Obviously,
such state is a partner of $X(5568)$ under the $SU(3)$ flavor symmetry, and their masses are close to each other.
But the threshold of $ud\bar{s}\bar{b}$ is $BK$, $270$ MeV higher than the threshold $B_{s}\pi$ of $X(5568)$
with $us\bar{d}\bar{b}$. So there is large mass region for this $ud\bar{s}\bar{b}$ state below threshold and
being stable. Besides, Ref.~\cite{YuFS} pointed out that if the lowest-lying $ud\bar{s}\bar{b}$ state exists
below threshold, it can be definitely observed via the weak decay mode $J/\psi K^{-}K^{-}\pi^{+}$, with the
expectation of hundreds of events in the current LHCb data sample but rejecting backgrounds due to its long
lifetime. Therefore, the $ud\bar{s}\bar{b}$ state would be a more promising detectable tetraquark state.
Ref.~\cite{ChenXY2} investigated such state composed of $ud\bar{s}\bar{b}$ within the chiral quark model,
and found the bound state with $IJ^{P}=00^{+}$ was possible. Liu $et~al.$ also proposed several partner states
of $X(5568)$ and estimated the mass difference of these partner states based on the color-magnetic
interaction~\cite{LiuYR}, which can provide valuable information on the future experimental search of these states.

It is generally known that quantum chromodynamics (QCD) is the fundamental theory of the strong interaction.
Understanding the low-energy behavior of QCD and the nature of the strong interacting matter, however, remains
a challenge due to the complexity of QCD. Lattice QCD has provided numerical results describing quark confinement
between two static colorful quarks, a preliminary picture of the QCD vacuum and the internal structure of hadrons
in addition to a phase transition of strongly interacting matter. But a satisfying description of multiquark
system is out of reach of the present calculation. The QCD-inspired models, incorporating the properties of
low-energy QCD: color confinement and chiral symmetry breaking, are also powerful tools to obtain physical
insights into many phenomena of the hadronic world.
Among many phenomenological models, the quark delocalization color screening model (QDCSM), which was developed
in the 1990s with the aim of explaining the similarities between nuclear (hadronic clusters
of quarks) and molecular forces~\cite{QDCSM0}, has been quite successful in reproducing the energies of
the baryon ground states, the properties of deuteron, the nucleon-nucleon ($NN$) and the hyperon-nucleon ($YN$)
interactions~\cite{QDCSM1}. Recently, this model has been used to study the pentaquarks with
hidden-strange~\cite{HuangHX_Ps}, hidden-charm and hidden-bottom~\cite{HuangHX_Pc}. Therefore,
it is interesting to extend this model to the tetraquark system. In present work, the tetraquark state
$X(5568)$ with quark contents $us\bar{d}\bar{b}$ and its partner state with $ud\bar{s}\bar{b}$ are
investigated. Besides, two structures, meson-meson and diquark-antidiquark, are considered in this work.

The structure of this paper is as follows. A brief introduction of the quark model and wave functions is
given in section II. Section III is devoted to the numerical
results and discussions. The summary is shown in the last section.

\section{MODEL AND WAVE FUNCTIONS}


QDCSM has been described in detail in the literatures~\cite{QDCSM0,QDCSM1}. Here, we
just present the salient features of the model. The Hamiltonian for the tetraquark states is shown below:
\begin{widetext}
\begin{eqnarray}
H & = & \sum_{i=1}^4\left(m_i+\frac{p_i^2}{2m_i}\right)-T_{CM} +\sum_{j>i=1}^4
\left(V^{CON}_{ij}+V^{OGE}_{ij}+V^{OBE}_{ij} \right), \\
V_{ij}^{CON} & = & \left \{
\begin{array}{ll}
-a_{c}\boldsymbol{\mathbf{\lambda}}^c_{i}\cdot
\boldsymbol{\mathbf{ \lambda}}^c_{j}~(r_{ij}^2+ a^{0}_{ij}), &
  \mbox{if \textit{i},\textit{j} in the same baryon orbit} \\
-a_{c}\boldsymbol{\mathbf{\lambda}}^c_{i}\cdot
\boldsymbol{\mathbf{
\lambda}}^c_{j}~(\frac{1-e^{-\mu_{ij}\mathbf{r}_{ij}^2}}{\mu_{ij}}+
a^{0}_{ij}), & \mbox{otherwise}
\end{array}
\right.\label{QDCSM-vc} \\
V^{OGE}_{ij} & = & \frac{1}{4}\alpha_s \boldsymbol{\lambda}^{c}_i \cdot
\boldsymbol{\lambda}^{c}_j
\left[\frac{1}{r_{ij}}-\frac{\pi}{2}\delta(\boldsymbol{r}_{ij})(\frac{1}{m^2_i}+\frac{1}{m^2_j}
+\frac{4\boldsymbol{\sigma}_i\cdot\boldsymbol{\sigma}_j}{3m_im_j})-\frac{3}{4m_im_jr^3_{ij}}
S_{ij}\right] \label{sala-vG} \\
V^{OBE}_{ij} & = & V_{\pi}( \boldsymbol{r}_{ij})\sum_{a=1}^3\lambda
_{i}^{a}\cdot \lambda
_{j}^{a}+V_{K}(\boldsymbol{r}_{ij})\sum_{a=4}^7\lambda
_{i}^{a}\cdot \lambda _{j}^{a}
+V_{\eta}(\boldsymbol{r}_{ij})\left[\left(\lambda _{i}^{8}\cdot
\lambda _{j}^{8}\right)\cos\theta_P-(\lambda _{i}^{0}\cdot
\lambda_{j}^{0}) \sin\theta_P\right] \label{sala-Vchi1} \\
V_{\chi}(\boldsymbol{r}_{ij}) & = & {\frac{g_{ch}^{2}}{{4\pi
}}}{\frac{m_{\chi}^{2}}{{\
12m_{i}m_{j}}}}{\frac{\Lambda _{\chi}^{2}}{{\Lambda _{\chi}^{2}-m_{\chi}^{2}}}}%
m_{\chi} \left\{(\boldsymbol{\sigma}_{i}\cdot
\boldsymbol{\sigma}_{j})
\left[ Y(m_{\chi}\,r_{ij})-{\frac{\Lambda_{\chi}^{3}}{m_{\chi}^{3}}}%
Y(\Lambda _{\chi}\,r_{ij})\right] \right.\nonumber \\
&& \left. +\left[H(m_{\chi}
r_{ij})-\frac{\Lambda_{\chi}^3}{m_{\chi}^3}
H(\Lambda_{\chi} r_{ij})\right] S_{ij} \right\}, ~~~~~~\chi=\pi, K, \eta, \\
S_{ij}&=&\left\{ 3\frac{(\boldsymbol{\sigma}_i
\cdot\boldsymbol{r}_{ij}) (\boldsymbol{\sigma}_j\cdot
\boldsymbol{r}_{ij})}{r_{ij}^2}-\boldsymbol{\sigma}_i \cdot
\boldsymbol{\sigma}_j\right\},\\
H(x)&=&(1+3/x+3/x^{2})Y(x),~~~~~~
 Y(x) =e^{-x}/x. \label{sala-vchi2}
\end{eqnarray}
\end{widetext}
Where $S_{ij}$ is quark tensor operator; $Y(x)$ and $H(x)$ are standard Yukawa functions; $T_c$ is the kinetic
energy of the center of mass; $\alpha_{s}$ is the quark-gluon coupling constant; $g_{ch}$ is the coupling constant
for chiral field, which is determined from the $NN\pi$ coupling constant through
\begin{equation}
\frac{g_{ch}^{2}}{4\pi }=\left( \frac{3}{5}\right) ^{2}{\frac{g_{\pi NN}^{2}%
}{{4\pi }}}{\frac{m_{u,d}^{2}}{m_{N}^{2}}}\label{gch}.
\end{equation}
The other symbols in the above expressions have their usual meanings. All model parameters are determined by
fitting the meson spectrum we used in this work and shown in Table~\ref{parameters}. The calculated masses of
the mesons in comparison with experimental values are shown in Table~\ref{mass}.
Besides, a phenomenological color screening confinement potential is used here, and $\mu_{ij}$ is the color
screening parameter, which is determined by fitting the deuteron properties, $NN$ scattering phase shifts,
$N\Lambda$ and $N\Sigma$ scattering phase shifts, respectively, with
$\mu_{uu}=0.45~$fm$^{-2}$, $\mu_{us}=0.19~$fm$^{-2}$ and $\mu_{ss}=0.08~$fm$^{-2}$, satisfying the relation,
$\mu_{us}^{2}=\mu_{uu}\mu_{ss}$~\cite{HuangHX3}. When
extending to the heavy bottom quark case, there is no experimental
data available, so we take it as a adjustable parameter $\mu_{bb}=0.001 \sim 0.0001~$fm$^{-2}$. We find the results
are insensitive to the value of $\mu_{bb}$. So in the present work, we take $\mu_{bb}=0.001~$fm$^{-2}$.

\begin{table}[ht]
\caption{\label{parameters}Model parameters:
$m_{\pi}=0.7$ fm$^{-1}$, $m_{ k}=2.51$ fm$^{-1}$,
$m_{\eta}=2.77$ fm$^{-1}$, $m_{\sigma}=3.42$ fm$^{-1}$,
$m_{a_{0}}=m_{\kappa}=m_{f_{0}}=4.97$ fm$^{-1}$,
$\Lambda_{\pi}=\Lambda_{\sigma}=4.2$ fm$^{-1}$, $\Lambda_{K}=\Lambda_{\eta}=
\Lambda_{a_{0}}=\Lambda_{\kappa}=\Lambda_{f_{0}}=5.2$ fm$^{-1}$,
$g_{ch}^2/(4\pi)$=0.54, $\theta_p$=$-15^{0}$. }
\begin{tabular}{ccccccccc} \hline\hline
$b$ & ~~~$m_{u}$~~~~ & ~~~$m_{d}$~~~ & ~~~$m_{s}$~~~ & ~~~$m_{b}$~~~~   \\
(fm) & (MeV) & (MeV) & (MeV) & (MeV)    \\ \hline
0.518  & 313 & 313 &  470  &    4500  \\ \hline
$ a_c$ &  $a^{0}_{uu}$ &  $a^{0}_{us}$ & $a^{0}_{ub}$ & $a^{0}_{sb}$  \\
(MeV\,fm$^{-2}$) & (fm$^{2}$) & (fm$^{2}$) & (fm$^{2}$) & (fm$^{2}$)  \\ \hline
  58.03 & -0.733 & -0.309 & 1.701 & 1.808 \\ \hline
 $\alpha_{s_{uu}}$ &  $\alpha_{s_{us}}$ & $\alpha_{s_{ub}}$ & $\alpha_{s_{sb}}$  \\ \hline
  1.50 & 1.46 & 1.41 & 1.40 \\
 \hline\hline
\end{tabular}
\end{table}

\begin{table}[ht]
\caption{The masses (in MeV) of the mesons obtained from QDCSM. Experimental values are taken
from the Particle Data Group (PDG)~\cite{PDG}.}
\begin{tabular}{lccccccccc}
\hline \hline
 ~~~~~Meson~~~~~& ~~~~~$M_{theo}$~~~~~ & ~~~~~$M_{exp}$~~~~~ \\ \hline
 ~~~~~~~~$\pi$     & 140 & 140 \\
 ~~~~~~~~$\rho$    & 772 & 770 \\
 ~~~~~~~~$K$       & 495 & 495 \\
 ~~~~~~~~$K^{*}$   & 892 & 892 \\
 ~~~~~~~~$B$       & 5280 & 5280 \\
 ~~~~~~~~$B^{*}$   & 5319 & 5325 \\
 ~~~~~~~~$B_{s}$   & 5367 & 5367 \\
 ~~~~~~~~$B^{*}_{s}$  & 5393 & 5415 \\ \hline\hline
\end{tabular}
\label{mass}
\end{table}

The quark delocalization in QDCSM is realized by specifying the single particle orbital wave function
of QDCSM as a linear combination of left and right Gaussians, the single particle
orbital wave functions used in the ordinary quark cluster model,
\begin{eqnarray}
\psi_{\alpha}(\mathbf{s}_i ,\epsilon) & = & \left(
\phi_{\alpha}(\mathbf{s}_i)
+ \epsilon \phi_{\alpha}(-\mathbf{s}_i)\right) /N(\epsilon), \nonumber \\
\psi_{\beta}(-\mathbf{s}_i ,\epsilon) & = &
\left(\phi_{\beta}(-\mathbf{s}_i)
+ \epsilon \phi_{\beta}(\mathbf{s}_i)\right) /N(\epsilon), \nonumber \\
N(\epsilon) & = & \sqrt{1+\epsilon^2+2\epsilon e^{-s_i^2/4b^2}}. \label{1q} \\
\phi_{\alpha}(\mathbf{s}_i) & = & \left( \frac{1}{\pi b^2}
\right)^{3/4}
   e^{-\frac{1}{2b^2} (\mathbf{r}_{\alpha} - \mathbf{s}_i/2)^2} \nonumber \\
\phi_{\beta}(-\mathbf{s}_i) & = & \left( \frac{1}{\pi b^2}
\right)^{3/4}
   e^{-\frac{1}{2b^2} (\mathbf{r}_{\beta} + \mathbf{s}_i/2)^2}. \nonumber
\end{eqnarray}
Here $\mathbf{s}_i$, $i=1,2,...,n$ are the generating coordinates,
which are introduced to expand the relative motion
wavefunction~\cite{QDCSM1}. The mixing parameter
$\epsilon(\mathbf{s}_i)$ is not an adjusted one but determined
variationally by the dynamics of the multi-quark system itself.
In this way, the multi-quark
system chooses its favorable configuration in the interacting process. This mechanism has been
used to explain the cross-over transition between hadron phase and quark-gluon plasma phase~\cite{Xu}.


In this work, the resonating group method (RGM)~\cite{RGM}, a well-established method for studying
a bound-state or a scattering problem, is used to calculate the energy of all these states. The wave
function of the four-quark system is of the form
\begin{equation}
\Psi = {\cal A } \left[[\psi^{L}\psi^{\sigma}]_{JM}\psi^{f}\psi^{c}\right].
\end{equation}
where $\psi^{L}$, $\psi^{\sigma}$, $\psi^{f}$, and $\psi^{c}$ are the orbital, spin, flavor and color
wave functions, respectively, which are given below. The symbol ${\cal A }$ is the anti-symmetrization operator.
For the meson-meson structure, ${\cal A }$ is defined as
\begin{equation}
{\cal A } = 1-P_{13}.
\end{equation}
where 1 and 3 stand for the quarks in two meson clusters respectively; for the diquark-antidiquark structure,
${\cal A }=1$.

The orbital wave function is in the form of
\begin{equation}
\psi^{L} = {\psi}_{1}(\boldsymbol{R}_{1}){\psi}_{2}(\boldsymbol{R}_{2})\chi_{L}(\boldsymbol{R}).
\end{equation}
where $\boldsymbol{R}_{1}$ and $\boldsymbol{R}_{2}$ are the internal coordinates for the cluster 1 and cluster 2.
$\boldsymbol{R} = \boldsymbol{R}_{1}-\boldsymbol{R}_{2}$ is the relative coordinate between the two clusters 1 and 2.
The ${\psi}_{1}$ and ${\psi}_{2}$ are the internal cluster orbital wave functions of the clusters 1 and 2,
and $\chi_{L}(\boldsymbol{R})$ is the relative motion wave function between two clusters,
which is expanded by gaussian bases
\begin{eqnarray}
& & \chi_{L}(\boldsymbol{R}) = \frac{1}{\sqrt{4\pi}}(\frac{3}{2\pi b^2}) \sum_{i=1}^{n} C_{i}  \nonumber \\
&& ~~~~\times  \int \exp\left[-\frac{3}{4b^2}(\boldsymbol{R}-\boldsymbol{s}_{i})^{2}\right] Y_{LM}(\hat{\boldsymbol{s}_{i}})d\hat{\boldsymbol{s}_{i}}. ~~~~~
\end{eqnarray}
where $\boldsymbol{s}_{i}$ is called the generate coordinate, $n$ is the number of the gaussian bases,
which is determined by the stability of the results. By doing this, the integro-differential equation
of RGM can be reduced to an algebraic equation, generalized eigen-equation. Then the energy of the system
can be obtained by solving this generalized eigen-equation. The details of solving the RGM equation can be found
in Ref.~\cite{RGM}. In our calculation, the maximum generating coordinate $s_n$ is fixed by the stability
of the results. The calculated results are stable when the distance between the two clusters is larger than 6 fm.
To keep the dimensions of matrix manageably small, the two clusters' separation is taken to be less than 6 fm.

The flavor, spin, and color wave functions are constructed in two steps. First constructing the wave functions
for clusters 1 and 2, then coupling the two wave functions of two clusters to form the wave function
for tetraquark system.
For the meson-meson structure, as the first step, we give the wave functions of the meson cluster.
The flavor wave functions of the meson cluster are shown below.
\begin{eqnarray}
\chi^{1}_{I_{11}} &=& u\bar{d},~~~~\chi^{2}_{I_{\frac{1}{2}\frac{1}{2}}} = s\bar{d},~~~~\chi^{3}_{I_{\frac{1}{2}\frac{1}{2}}} = u\bar{b},~~~~\chi^{4}_{I_{00}} = s\bar{b},  \nonumber \\
\chi^{5}_{I_{\frac{1}{2}\frac{1}{2}}} &=& u\bar{s},~~~~\chi^{6}_{I_{\frac{1}{2}-\frac{1}{2}}} = d\bar{s},~~~~\chi^{7}_{I_{\frac{1}{2}-\frac{1}{2}}} = d\bar{b}.
\end{eqnarray}
where the superscript of the $\chi$ is the index of the flavor wave function for a meson, and the subscript
stands for the isospin  $I$ and the third component $I_{z}$. The spin wave functions of the meson cluster are:
\begin{eqnarray}
\chi^{1}_{\sigma_{11}} &=& \alpha\alpha,~~~~\chi^{2}_{\sigma_{10}} = \sqrt{\frac{1}{2}}(\alpha\beta+\beta\alpha), \nonumber \\
~~~~\chi^{3}_{\sigma_{1-1}} &=& \beta\beta,~~~~\chi^{4}_{\sigma_{00}} = \sqrt{\frac{1}{2}}(\alpha\beta-\beta\alpha).
\end{eqnarray}
and the color wave function of a meson is:
\begin{eqnarray}
\chi^{1}_{[111]} &=& \sqrt{\frac{1}{3}}(r\bar{r}+g\bar{g}+b\bar{b}).
\end{eqnarray}
Then, the wave functions for the four-quark system with the meson-meson structure can be obtained by coupling
the wave functions of two meson clusters. Every part of wave functions are shown below. The flavor wave functions are:
\begin{eqnarray}
\psi^{f_{1}}_{11} &=& \chi^{4}_{I_{00}}\chi^{1}_{I_{11}},~~~~\psi^{f_{2}}_{11} = \chi^{3}_{I_{\frac{1}{2}\frac{1}{2}}}\chi^{2}_{I_{\frac{1}{2}\frac{1}{2}}},  \nonumber \\
\psi^{f_{3}}_{00} &=& \sqrt{\frac{1}{2}}\left[\chi^{5}_{I_{\frac{1}{2}\frac{1}{2}}}\chi^{7}_{I_{\frac{1}{2}-\frac{1}{2}}}-\chi^{7}_{I_{\frac{1}{2}-\frac{1}{2}}}\chi^{5}_{I_{\frac{1}{2}\frac{1}{2}}}\right],  \nonumber \\
\psi^{f_{4}}_{11} &=& \sqrt{\frac{1}{2}}\left[\chi^{5}_{I_{\frac{1}{2}\frac{1}{2}}}\chi^{7}_{I_{\frac{1}{2}-\frac{1}{2}}}+\chi^{7}_{I_{\frac{1}{2}-\frac{1}{2}}}\chi^{5}_{I_{\frac{1}{2}\frac{1}{2}}}\right].
\end{eqnarray}
The spin wave functions are:
\begin{eqnarray}
\psi^{\sigma_{1}}_{00} &=& \chi^{4}_{\sigma_{00}}\chi^{4}_{\sigma_{00}},\nonumber \\
\psi^{\sigma_{2}}_{00} &=& \sqrt{\frac{1}{3}}\left[\chi^{1}_{\sigma_{11}}\chi^{3}_{\sigma_{1-1}}-\chi^{2}_{\sigma_{10}}\chi^{2}_{\sigma_{10}}+\chi^{3}_{\sigma_{1-1}}\chi^{1}_{\sigma_{11}}\right],  \nonumber \\
\psi^{\sigma_{3}}_{11} &=& \chi^{4}_{\sigma_{00}}\chi^{1}_{\sigma_{11}},~~~~\psi^{\sigma_{4}}_{11} = \chi^{1}_{\sigma_{11}}\chi^{4}_{\sigma_{00}},  \nonumber \\
\psi^{\sigma_{5}}_{11} &=& \sqrt{\frac{1}{2}}\left[\chi^{1}_{\sigma_{11}}\chi^{2}_{\sigma_{10}}-\chi^{2}_{\sigma_{10}}\chi^{1}_{\sigma_{11}}\right].
\end{eqnarray}
The color wave function is:
\begin{eqnarray}
\psi^{c_{1}} &=& \chi^{1}_{[111]}\chi^{1}_{[111]}.
\end{eqnarray}
Finally, multiplying the wave functions $\psi^{L}$, $\psi^{\sigma}$, $\psi^{f}$, and $\psi^{c}$ according to
the definite quantum number of the system, we can acquire the total wave functions of the system.

For the diquark-antidiquark structure, the orbital and the spin wave functions are the same with those of
the meson-meson structure. For the flavor wave functions, we give the functions of the diquark and antidiquark
clusters firstly.
\begin{eqnarray}
\chi^{1}_{I_{10}} &=& \frac{1}{\sqrt{2}}(ud+du),~~~~\chi^{2}_{I_{00}} = \frac{1}{\sqrt{2}}(ud-du),  \nonumber \\
\chi^{3}_{I_{\frac{1}{2}\frac{1}{2}}} &=& \frac{1}{\sqrt{2}}(us+su),~~~~\chi^{4}_{I_{\frac{1}{2}\frac{1}{2}}} = \frac{1}{\sqrt{2}}(us-su),  \nonumber \\
\chi^{5}_{I_{\frac{1}{2}\frac{1}{2}}} &=& \bar{d}\bar{b},~~~~\chi^{6}_{I_{00}} = \bar{s}\bar{b}.
\end{eqnarray}

Then, the color wave functions of the diquark clusters are:
\begin{eqnarray}
\chi^{1}_{[2]} &=& rr,~~~\chi^{2}_{[2]} = \frac{1}{\sqrt{2}}(rg+gr),~~~\chi^{3}_{[2]} = gg,  \nonumber \\
\chi^{4}_{[2]} &=& \frac{1}{\sqrt{2}}(rb+br),~~~\chi^{5}_{[2]} = \frac{1}{\sqrt{2}}(gb+bg),~~~\chi^{6}_{[2]} = bb,  \nonumber \\
\chi^{7}_{[11]}&=& \frac{1}{\sqrt{2}}(rg-gr),~~~\chi^{8}_{[11]}= \frac{1}{\sqrt{2}}(rb-br), \nonumber \\
\chi^{9}_{[11]}&=& \frac{1}{\sqrt{2}}(gb-bg).
\end{eqnarray}
and the color wave functions of the antidiquark clusters are:
\begin{eqnarray}
\chi^{1}_{[22]} &=& \bar{r}\bar{r},~~~\chi^{2}_{[22]} = -\frac{1}{\sqrt{2}}(\bar{r}\bar{g}+\bar{g}\bar{r}),~~~~\chi^{3}_{[22]} = \bar{g}\bar{g},  \nonumber \\
\chi^{4}_{[22]} &=& \frac{1}{\sqrt{22}}(\bar{r}\bar{b}+\bar{b}\bar{r}),~\chi^{5}_{[22]} = -\frac{1}{\sqrt{2}}(\bar{g}\bar{b}+\bar{b}\bar{g}),~\chi^{6}_{[22]} = \bar{b}\bar{b},  \nonumber \\
\chi^{7}_{[211]}&=& \frac{1}{\sqrt{2}}(\bar{r}\bar{g}-\bar{g}\bar{r}),~~~~\chi^{8}_{[211]}= -\frac{1}{\sqrt{2}}(\bar{r}\bar{b}-\bar{b}\bar{r}), \nonumber \\
\chi^{9}_{[211]}&=& \frac{1}{\sqrt{2}}(\bar{g}\bar{b}-\bar{b}\bar{g}).
\end{eqnarray}

After that, the wave functions for the four-quark system with the diquark-antidiquark structure can be obtained
by coupling the wave functions of two clusters. Every part of wave functions are shown below. The flavor wave functions are:
\begin{eqnarray}
\psi^{f_{1}}_{11} &=& \chi^{3}_{I_{\frac{1}{2}\frac{1}{2}}}\chi^{5}_{I_{\frac{1}{2}\frac{1}{2}}},~~~~\psi^{f_{2}}_{11} = \chi^{4}_{I_{\frac{1}{2}\frac{1}{2}}}\chi^{5}_{I_{\frac{1}{2}\frac{1}{2}}},  \nonumber \\
\psi^{f_{3}}_{11} &=& \chi^{1}_{I_{11}}\chi^{6}_{I_{00}},~~~~\psi^{f_{4}}_{00} = \chi^{2}_{I_{00}}\chi^{6}_{I_{00}}.
\end{eqnarray}
The color wave functions are:
\begin{eqnarray}
\psi^{c_{1}} &=& \sqrt{\frac{1}{6}}[\chi^{1}_{[2]}\chi^{1}_{[22]}-\chi^{2}_{[2]}\chi^{2}_{[22]}+\chi^{3}_{[2]}\chi^{3}_{[22]}  \nonumber \\
&& +\chi^{4}_{[2]}\chi^{4}_{[22]}-\chi^{5}_{[2]}\chi^{5}_{[22]}+\chi^{6}_{[2]}\chi^{6}_{[22]}],  \nonumber \\
\psi^{c_{2}} &=& \sqrt{\frac{1}{3}}\left[\chi^{7}_{[11]}\chi^{7}_{[211]}-\chi^{8}_{[11]}\chi^{8}_{[211]}+\chi^{9}_{[11]}\chi^{9}_{[211]}\right].
\end{eqnarray}

Finally, we can acquire the total wave functions by substituting the wave functions of the orbital, the spin,
the flavor and the color parts into the Eq. (10) according to the given quantum number of the system.

\section{The results and discussions}

In present work, we investigate tetraquarks with quark components: $us\bar{d}\bar{b}$ and $ud\bar{s}\bar{b}$
in two structures, meson-meson and diquark-antidiquark. The quantum numbers of the tetraquarks we study here are
$I=0,~1$, $J=0,~1$ and the parity is $P=+$. The orbital angular momenta are set to zero because we are interested
in the ground states. To check whether or not there is any bound state in such tetraquark system, we do a dynamic
bound-state calculation. Both the single-channel and channel-coupling calculations are carried out in this work.
All the general features of the calculated results are as follows.

\subsection{Tetraquarks composed of $us\bar{d}\bar{b}$}
For tetraquarks composed of $us\bar{d}\bar{b}$, the isospin is $I=1$. The energies of the states with $J=0,1$
are calculated and the results are listed in Table~\ref{bound1} and \ref{bound2}. In the tables, the second column
gives the index of the wave functions of each channel. The columns headed with $E_{sc}$ and $E_{cc}$ represent the
energies of the single-channel and channel-coupling calculation respectively. For meson-meson structure, there are
two additional columns, the column headed with ``Channel" denotes the physical contents of the channel and the coulmn
headed with $E_{th}$ denotes the theoretical threshold of the channel. From the Table~\ref{bound1}, we can see that
the energies of every single channel approach to the corresponding theoretical threshold. The channel-coupling cannot
help too much. Energies are still above the threshold of the lowest channel ($B^{0}_{s}\pi^{+}$ for $IJ=10$ and
$B^{*0}_{s}\pi^{+}$ for $IJ=11$), which indicates that no bound $us\bar{d}\bar{b}$ state with meson-meson structure
is formed in our quark model calculation.

\begin{table}
\caption{The energies (in MeV) of the meson-meson structure for tetraquarks $us\bar{d}\bar{b}$.}
\begin{tabular}{lcccccccccc}
\hline \hline
  & ~$[\psi^{f_{i}}\psi^{\sigma_{j}}\psi^{c_{k}}]$~ & ~Channel~ & ~$E_{th}$~~ & ~~$E_{sc}$~~ & ~~$E_{cc}$~~  \\ \hline
 ~~$IJ=10$~~ & $[\psi^{f_{1}}\psi^{\sigma_{1}}\psi^{c_{1}}]$ & $B^{0}_{s}\pi^{+}$ & $5506.9$ & $5514.0$ & $5513.1$ \\
             & $[\psi^{f_{1}}\psi^{\sigma_{2}}\psi^{c_{1}}]$ & $B^{*0}_{s}\rho^{+}$ & $6165.4$ & $6169.2$ &  \\
             & $[\psi^{f_{2}}\psi^{\sigma_{1}}\psi^{c_{1}}]$ & $B^{+}\bar{K}^{0}$ & $5774.9$ & $5782.6$ &  \\
             & $[\psi^{f_{2}}\psi^{\sigma_{2}}\psi^{c_{1}}]$ & $B^{*+}\bar{K}^{*0}$ & $6212.6$ & $6217.5$ &  \\ \hline
 ~~$IJ=11$~~ & $[\psi^{f_{1}}\psi^{\sigma_{3}}\psi^{c_{1}}]$ & $B^{0}_{s}\rho^{+}$ & $6139.4$ & $6143.9$ & $5539.3$ \\
             & $[\psi^{f_{1}}\psi^{\sigma_{4}}\psi^{c_{1}}]$ & $B^{*0}_{s}\pi^{+}$ & $5532.9$ & $5539.7$ &  \\
             & $[\psi^{f_{2}}\psi^{\sigma_{3}}\psi^{c_{1}}]$ & $B^{+}\bar{K}^{*0}$ & $6172.3$ & $6179.1$ &  \\
             & $[\psi^{f_{2}}\psi^{\sigma_{4}}\psi^{c_{1}}]$ & $B^{*+}\bar{K}^{0}$ & $5814.2$ & $5821.5$ &  \\
 \hline\hline
\end{tabular}
\label{bound1}
\end{table}

\begin{table}
\caption{The energies (in MeV) of the diquark-antidiquark structure for tetraquarks $us\bar{d}\bar{b}$.}
\begin{tabular}{lcccccccccc}
\hline \hline
  & ~$[\psi^{f_{i}}\psi^{\sigma_{j}}\psi^{c_{k}}]$~ &  ~~~~$E_{sc}$~~~~ & ~~~~$E_{cc}$~~~~  \\ \hline
 ~~$IJ=10$~~ & $[\psi^{f_{1}}\psi^{\sigma_{1}}\psi^{c_{1}}]$  & $6283.1$ & $5551.8$ \\
             & $[\psi^{f_{1}}\psi^{\sigma_{2}}\psi^{c_{2}}]$  & $6186.9$ &  \\
             & $[\psi^{f_{2}}\psi^{\sigma_{1}}\psi^{c_{2}}]$  & $6096.1$ &  \\
             & $[\psi^{f_{2}}\psi^{\sigma_{2}}\psi^{c_{1}}]$  & $5846.6$ &  \\ \hline
 ~~$IJ=11$~~ & $[\psi^{f_{1}}\psi^{\sigma_{4}}\psi^{c_{2}}]$  & $6308.8$ & $5613.4$ \\
             & $[\psi^{f_{1}}\psi^{\sigma_{5}}\psi^{c_{2}}]$  & $6261.9$ &  \\
             & $[\psi^{f_{1}}\psi^{\sigma_{3}}\psi^{c_{1}}]$  & $6276.0$ &  \\
             & $[\psi^{f_{2}}\psi^{\sigma_{4}}\psi^{c_{1}}]$  & $6216.5$ &  \\
             & $[\psi^{f_{2}}\psi^{\sigma_{5}}\psi^{c_{1}}]$  & $6059.4$ &  \\
             & $[\psi^{f_{2}}\psi^{\sigma_{3}}\psi^{c_{2}}]$  & $6118.7$ &  \\
 \hline\hline
\end{tabular}
\label{bound2}
\end{table}

With regard to the diquark-antidiquark structure, the energies are listed in Table~\ref{bound2}. The channels
with different flavor-spin-color configurations have different energies and the coupling of them is rather
stronger than that of the meson-meson structure. However, the energy of the $IJ=10$ state is still higher than
the theoretical threshold of the lowest channel $B^{0}_{s}\pi^{+}$, $5506.9$ MeV. Similarly, the energy of the
$IJ=11$ state is higher than the theoretical threshold of the lowest channel $B^{*0}_{s}\pi^{+}$, $5532.9$ MeV.
Thus, there is no bound state with diquark-antidiquark structure in the present calculation.

Nevertheless, the colorful subclusters diquark ($qq$) and antidiquark ($\bar{q}\bar{q}$) cannot fall apart because
of the color confinement, so there may be a resonance state with diquark-antidiquark structure. To check the
possibility, we perform an adiabatic calculation of energy for both the $IJ=10$ and $IJ=11$ states. The results are
shown in Fig. 1, where the horizontal axis $S$ is the distance between two subclusters and the vertical axis stands
for the energy of the system at the corresponding distance $S$. It is obvious in Fig. 1 that the energy of both the
$IJ=10$ and $IJ=11$ states is increasing when the two subclusters fall apart, which indicates that the two subclusters
tend to clump together. In other words, the odds are the same for the states being meson-meson structure,
diquark-antidiquark structure or other structures. As mentioned above, the energy of the state is higher than the
theoretical threshold of the lowest channel, so neither the state of $IJ=10$ nor the state of $IJ=11$ is a resonance
state in QDCSM.

Therefore, the $X(5568)$ cannot be explained as a molecular state or a diquark-antidiquark resonance of
$us\bar{d}\bar{b}$ in the present calculation. Our results are consistent with the analysis of Ref.~\cite{ChenXY}
and Ref.~\cite{Burns}. In Ref~\cite{ChenXY}, the four-quark system $us\bar{d}\bar{b}$ with both meson-meson structure and diquark-antidiquark structure was studied in the framework of the chiral quark model by using the Gaussian
expansion method, and no candidate of $X(5568)$ was found. In Ref.~\cite{Burns}, Burns and Swanson explored a lot of
possible explanations of the $X(5568)$ signal, a tetraquark, a hadronic molecule or a threshold effect and
found that none of them can be a candidate of the observed state.

\begin{figure}[ht]
\begin{center}
\epsfxsize=3.0in \epsfbox{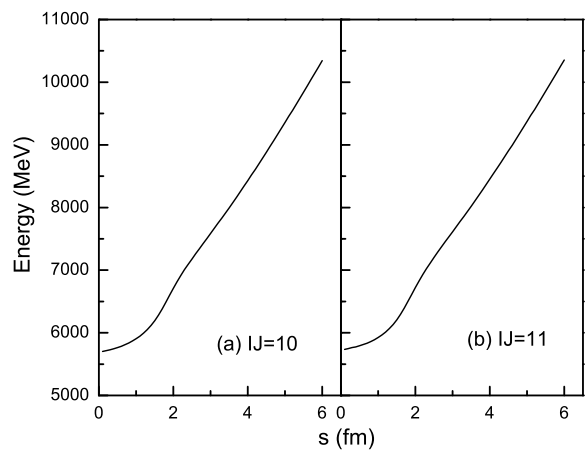} \vspace{-0.1in}

\caption{The energy as a function of the distance between the diquark ($qq$) and antidiquark ($\bar{q}\bar{q}$) for the $IJ=10$ and $IJ=11$ states of $us\bar{d}\bar{b}$.}
\end{center}
\end{figure}

\subsection{Tetraquarks composed of $ud\bar{s}\bar{b}$}
\begin{table}
\caption{The energies (in MeV) of the meson-meson structure for tetraquarks $ud\bar{s}\bar{b}$.}
\begin{tabular}{lcccccccccc}
\hline \hline
  & ~$[\psi^{f_{i}}\psi^{\sigma_{j}}\psi^{c_{k}}]$~ & ~Channel~ & ~$E_{th}$~~ & ~~$E_{sc}$~~ & ~~$E_{cc}$~~  \\ \hline
 ~~$IJ=00$~~ & $[\psi^{f_{3}}\psi^{\sigma_{1}}\psi^{c_{1}}]$ & $B^{0}K^{+}$ & $5774.9$ & $5781.4$ & $5779.9$ \\
             & $[\psi^{f_{3}}\psi^{\sigma_{2}}\psi^{c_{1}}]$ & $B^{*0}K^{*+}$ & $6212.6$ & $6218.1$ &  \\ \hline
 ~~$IJ=01$~~ & $[\psi^{f_{3}}\psi^{\sigma_{3}}\psi^{c_{1}}]$ & $B^{0}K^{*+}$ & $6172.3$ & $6176.2$ & $5813.1$ \\
             & $[\psi^{f_{3}}\psi^{\sigma_{4}}\psi^{c_{1}}]$ & $B^{*0}K^{+}$ & $5814.2$ & $5820.5$ &  \\
             & $[\psi^{f_{3}}\psi^{\sigma_{5}}\psi^{c_{1}}]$ & $B^{*0}K^{*+}$ & $6212.6$ & $6216.1$ &  \\\hline
 ~~$IJ=10$~~ & $[\psi^{f_{4}}\psi^{\sigma_{1}}\psi^{c_{1}}]$ & $B^{0}K^{+}$ & $5774.9$ & $5783.2$ & $5783.0$ \\
             & $[\psi^{f_{4}}\psi^{\sigma_{2}}\psi^{c_{1}}]$ & $B^{*0}K^{*+}$ & $6212.6$ & $6218.1$ &  \\ \hline
 ~~$IJ=11$~~ & $[\psi^{f_{4}}\psi^{\sigma_{3}}\psi^{c_{1}}]$ & $B^{0}K^{*+}$ & $6172.3$ & $6180.3$ & $5821.5$ \\
             & $[\psi^{f_{4}}\psi^{\sigma_{4}}\psi^{c_{1}}]$ & $B^{*0}K^{+}$ & $5814.2$ & $5822.0$ &  \\
             & $[\psi^{f_{4}}\psi^{\sigma_{5}}\psi^{c_{1}}]$ & $B^{*0}K^{*+}$ & $6212.6$ & $6219.1$ &  \\\hline
 \hline\hline
\end{tabular}
\label{bound3}
\end{table}

\begin{table}
\caption{The energies (in MeV) of the diquark-antidiquark structure for tetraquarks $ud\bar{s}\bar{b}$.}
\begin{tabular}{lcccccccccc}
\hline \hline
  & ~$[\psi^{f_{i}}\psi^{\sigma_{j}}\psi^{c_{k}}]$~ &  ~~~~$E_{sc}$~~~~ & ~~~~$E_{cc}$~~~~  \\ \hline
 ~~$IJ=00$~~ & $[\psi^{f_{1}}\psi^{\sigma_{1}}\psi^{c_{1}}]$  & $5867.2$ & $5701.1$ \\
             & $[\psi^{f_{1}}\psi^{\sigma_{2}}\psi^{c_{2}}]$  & $6058.2$ &  \\ \hline
 ~~$IJ=01$~~ & $[\psi^{f_{1}}\psi^{\sigma_{1}}\psi^{c_{1}}]$  & $6334.7$ & $5756.3$ \\
             & $[\psi^{f_{2}}\psi^{\sigma_{1}}\psi^{c_{2}}]$  & $6213.3$ &  \\
             & $[\psi^{f_{2}}\psi^{\sigma_{2}}\psi^{c_{1}}]$  & $5881.4$ &  \\ \hline
 ~~$IJ=10$~~ & $[\psi^{f_{1}}\psi^{\sigma_{1}}\psi^{c_{1}}]$  & $6452.1$ & $6103.8$ \\
             & $[\psi^{f_{1}}\psi^{\sigma_{2}}\psi^{c_{2}}]$  & $6200.8$ &  \\ \hline
 ~~$IJ=11$~~ & $[\psi^{f_{1}}\psi^{\sigma_{1}}\psi^{c_{1}}]$  & $6289.1$ & $6130.1$ \\
             & $[\psi^{f_{2}}\psi^{\sigma_{1}}\psi^{c_{2}}]$  & $6253.4$ &  \\
             & $[\psi^{f_{2}}\psi^{\sigma_{2}}\psi^{c_{1}}]$  & $6447.9$ &  \\
 \hline\hline
\end{tabular}
\label{bound4}
\end{table}

For tetraquarks composed of $ud\bar{s}\bar{b}$, four states with the quantum numbers $IJ=00, 01, 10$
and $11$ are studied. The energies of the meson-meson structure and the diquark-antidiquark structure are listed in
Tables~\ref{bound3} and \ref{bound4}, respectively. For the meson-meson structure, the results are similar to
that of the tetraquarks of $us\bar{d}\bar{b}$. Table~\ref{bound3} shows that the energies of every single channel
are above the corresponding theoretical threshold. The effect of channel-coupling is very small except for the
$IJ=01$ state. For the states with $IJ=00$, $IJ=10$, and $IJ=11$, all energies are above the threshold of the
lowest channel ($B^{0}K^{+}$ for $IJ=00$, $B^{0}K^{+}$ for $IJ=10$, and $B^{*0}K^{+}$ for $IJ=11$) even by the
channel-coupling calculation. While for the state with $IJ=01$, the energy is about $1.0$ MeV lower than the
threshold of the lowest channel $B^{*0}K^{+}$ after channel-coupling. However, the binding energy is not very large,
so there maybe a weak molecular bound state of $ud\bar{s}\bar{b}$ with quantum numbers of $IJ=01$, and the mass
of this state is about $5813$ MeV.

For the diquark-antidiquark structure, all the possible channels are shown in Table~\ref{bound4}. One can see that
the energy of each single channel is higher than the theoretical threshold of the corresponding channel, which are
shown in Table~\ref{bound3}. Although the effect of the channel-coupling is much stronger than that of the meson-meson
structure, the energy of the $IJ=10$ and $IJ=11$ states are still above the theoretical threshold of the corresponding
channel. So there is no any bound states for the $IJ=10$ or $IJ=11$ state. In order to check if there is any resonance
state, we also perform the adiabatic calculation of energy for both the $IJ=10$ and $IJ=11$ states. The results are shown
in Fig. 2. The case is similar to the tetraquarks composed of $us\bar{d}\bar{b}$. The energy of both the $IJ=10$ and $IJ=11$
states is increasing when the two subclusters diquark ($qq$) and antidiquark ($\bar{q}\bar{q}$) fall apart, which indicates
that the two subclusters tend to clump together. So there is no resonance state with quantum numbers $IJ=10$ and $IJ=11$.

However, things are different for the $IJ=00$ state and the $IJ=01$ state. The energy of the $IJ=00$ state is about
$5701$ MeV, $74$ MeV lower than the theoretical threshold of the $B^{0}K^{+}$, which indicates that the $IJ=00$ state
of the diquark-antidiquark structure can be a bound state. Ref.~\cite{ChenXY2} also found that the bound state with
$IJ=00$ was possible. Meanwhile, the energy of the $IJ=01$ state is $58$ MeV lower than the theoretical threshold
of the $B^{*0}K^{+}$, so the $IJ=01$ state is also bound here. Thus, both the $IJ=00$ state and the $IJ=01$ state of
diquark-antidiquark structure can form bound states. 

By comparing with the results of the meson-meson structure, we note that the energy of the $IJ=01$ state of
the diquark-antidiquark structure is about $5756$ MeV, which is much lower than that of the meson-meson structure
shown in Table~\ref{bound3}. This shows that the $IJ=01$ state prefers to be a bound state of the diquark-antidiquark
structure. Moreover, the $IJ=00$ state of the diquark-antidiquark structure is easier to form the bound state than
the one of the meson-meson structure. All these indicate that the diquark-antidiquark configuration maybe a good
choice for some tetraquarks. Some work have been done to explain the exotic $XYZ$ states depending on the
diquark-antidiquark configuration. Ref.~\cite{Maiani} proposed the hypothesis that the diquarks and antidiquarks
in tetraquarks were separated by a potential barrier to explain the properties of exotic resonances such as $X$ and $Z$.
Ref.~\cite{Brodsky} presented a dynamical picture to explain the nature of some exotic $XYZ$ states based on a
diquark-antidiquark open-string configuration. The picture combined the advantages of diquark-based models, which
can accommodate much of the known $XYZ$ spectrum, with the experimental fact that such states are both relatively
narrow and are produced promptly. Thus both the $IJ=00$ and $IJ=01$ states of the diquark-antidiquark structure
we obtain here maybe the narrow resonance states. The study of the decay width of these states is our further work.

Contrasting with the tetraquarks composed of $us\bar{d}\bar{b}$, we find the tetraquarks composed of $ud\bar{s}\bar{b}$
is more likely to form bound state. The reasons are as follows. First, the diquark pair of two light quarks ($ud$)
or two heavier quarks ($sb$) is usually more stable than the one of two quarks with larger mass difference like
$us$ or $db$ pair. Our results show that the tetraquarks composed of $ud\bar{s}\bar{b}$ of the diquark-antidiquark
structure is most possible to form bound states, which just supports this point. Secondly, the lowest threshold of
$ud\bar{s}\bar{b}$ is $BK$, $270$ MeV higher than the threshold $B_{s}\pi$ of $us\bar{d}\bar{b}$. So there is large
mass region for this $ud\bar{s}\bar{b}$ state below threshold and being stable. This conclusion also supports the 
assumption of Ref.~\cite{YuFS}, which proposed such particle with the quark component of $ud\bar{s}\bar{b}$ 
(or its charge-conjugated one) as a partner of $X(5568)$ of $us\bar{d}\bar{b}$ under the $SU(3)$ flavor symmetry.

\begin{figure}[ht]
\begin{center}
\epsfxsize=3.0in \epsfbox{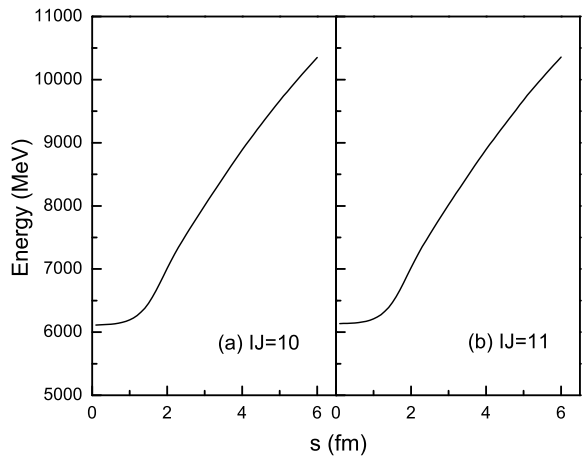} \vspace{-0.1in}

\caption{The energy as a function of the distance between the diquark ($qq$) and antidiquark ($\bar{q}\bar{q}$) 
for the $IJ=10$ and $IJ=11$ states of $ud\bar{s}\bar{b}$.}
\end{center}
\end{figure}

\section{Summary}
In summary, we investigate tetraquarks composed of $us\bar{d}\bar{b}$ and $ud\bar{s}\bar{b}$ in the framework of QDCSM. 
Two structures, meson-meson and diquark-antidiquark, are considered.
Our results show that there is no bound state or resonance state composed of $us\bar{d}\bar{b}$. The reported $X(5568)$ 
cannot be explained as a molecular state or a diquark-antidiquark resonance of $us\bar{d}\bar{b}$ in present calculation. 
In contrast, two bound states are obtained for the tetraquarks system composed of $ud\bar{s}\bar{b}$: an $IJ=00$ state 
with the mass of $5701$ MeV, and an $IJ=01$ state with the mass of $5756$ MeV, which maybe the better tetraquark candidates
with foure different flavors. These two bound states are of the diquark-antidiquark structure. For the system with 
$IJ=00$ and $IJ=01$, it is obvious that the state of the diquark-antidiquark structure is more likely to form bound 
state than that of the meson-meson structure, which indicates that the diquark-antidiquark configuration would be a 
good choice for the tetraquarks $ud\bar{s}\bar{b}$ with $IJ=00$ and $IJ=01$. During the calculation, we find that 
the effect of the channel-coupling in the diquark-antidiquark structure is much stronger than that in the meson-meson 
structure, and the channel-coupling plays an important role in forming bound states in the diquark-antidiquark structure.

Meanwhile, our results also show that the tetraquarks composed of $ud\bar{s}\bar{b}$ is more possible to form bound 
states than the one composed of $us\bar{d}\bar{b}$. Thus, if the $X(5568)$ does exist, the tetraquarks composed of 
$ud\bar{s}\bar{b}$ must be a more stable state. If the $X(5568)$ is proved to be nonexistent, it is still possible 
for the existence of such tetraquarks with $ud\bar{s}\bar{b}$ components. This conclusion is in accordance with 
the point of Ref.~\cite{YuFS}, which proposed the state composed of $ud\bar{s}\bar{b}$ (or its charge-conjugated one) 
as a partner of $X(5568)$ of $us\bar{d}\bar{b}$. Ref.~\cite{YuFS} also pointed out that if the lowest-lying 
$ud\bar{s}\bar{b}$ state exists below threshold, it can be definitely observed via the weak decay mode 
$J/\psi K^{-}K^{-}\pi^{+}$, with the expectation of hundreds of events in the current LHCb data sample but 
rejecting backgrounds due to its long lifetime. Therefore, the $ud\bar{s}\bar{b}$ state would be a promising 
detectable tetraquark state. We hope that experiments will help to discover these interesting tetraquark states.

\acknowledgments{This work is supported partly by the National Science Foundation
of China under Contract Nos. 11675080, 11775118 and 11535005, the Natural Science Foundation of
the Jiangsu Higher Education Institutions of China (Grant No. 16KJB140006).

\end{document}